\documentclass[prd,aps,showpacs,nofootinbib]{revtex4}%
\usepackage{amssymb}
\usepackage{graphicx}
\usepackage{amsmath}
\usepackage{amsfonts}%
\newcommand{\be}{\begin{equation}}
\newcommand{\ee}{\end{equation}}
\newcommand{\bq}{\begin{eqnarray}}
\newcommand{\eq}{\end{eqnarray}}
\begin{document}
\title{\textbf{Stationary solutions for the parity-even sector of the CPT-even and
Lorentz-covariance-violating term of the standard model extension}}
\author{Rodolfo Casana$^{a}$, Manoel M. Ferreira Jr$^{a}$, A. R. Gomes$^{b}$, Paulo R.
D. Pinheiro$^{a}$}
\affiliation{$^{a}$Departamento de F\'{\i}sica, Universidade Federal do Maranh\~{a}o
(UFMA), Campus Universit\'{a}rio do Bacanga, S\~{a}o Lu\'{\i}s - MA,
65085-580, Brasil}
\affiliation{$^{b}$Departamento de Ci\^{e}ncias Exatas, Centro Federal de Educa\c{c}\~{a}o
Tecnol\'{o}gica do Maranh\~{a}o, 65025-001 S\~{a}o Lu\'{\i}s, Maranh\~{a}o, Brazil}

\begin{abstract}
In this work, we focus on some properties of the parity-even sector of the
CPT-even electrodynamics of the standard model extension. We analyze how the
six non-birefringent terms belonging to this sector modify the static and
stationary classical solutions of the usual Maxwell theory. We observe that
the parity-even terms do not couple the electric and magnetic sectors\ (at
least in the stationary regime). The Green's method is used to obtain
solutions for the field strengths $E$ and $B$ at first order in the Lorentz-
covariance-violating parameters. Explicit solutions are attained for
point-like and spatially extended sources, for which a dipolar expansion is
achieved. Finally, it is presented an Earth-based experiment that can lead (in
principle) to an upper bound on the anisotropic coefficients as stringent as
$\left(  \widetilde{\kappa}_{e-}\right)  ^{ij}<2.9\times10^{-20}.$

\end{abstract}

\maketitle

\section{Introduction}

In recent years, investigations concerning Lorentz symmetry
violation have been undertaken mainly in the context of the standard
model extension (SME) developed by Colladay and Kostelecky
\cite{Colladay}, \cite{Coleman}, which incorporates
Lorentz-invariance-violating (LIV) terms in all sectors of the usual
standard model of the fundamental interactions. The abelian or
electromagnetic sector of the SME is composed of a CPT-even and a
CPT-odd part. The CPT-odd sector is represented by the
Carroll-Field-Jackiw term,
$\varepsilon_{\beta\alpha\rho\varphi}V^{\beta}A^{\alpha}F^{\rho\varphi},$
whose properties were first examined in Ref. \cite{Jackiw}. The
investigations on this electrodynamics have been performed in a
broad perspective, addressing aspects as diverse as the consistency
and quantization\ of the model \cite{Adam}, radiative corrections
\cite{Radiative}, classical solutions \cite{Classical}, Cerenkov
radiation \cite{Cerenkov}, cosmic background radiation \cite{CBR},
and other features \cite{General}. More recently, the CPT-even
sector, represented by the term
$W_{\alpha\nu\rho\varphi}F^{\alpha\nu}F^{\rho\varphi },$ has been
investigated as well, embracing the study of small deviations of the
Maxwell electrodynamics stemming from this term and some attempts of
imposing upper bounds on the LIV parameters \cite{Bailey},
\cite{Manojr}, \cite{Altschul},
\cite{KM1},\cite{KM2},\cite{Kob},\cite{Klink},\cite{Klink3}.

The Lagrangian density of the CPT-even electrodynamics of the Standard Model
Extension has the form
\begin{equation}
\mathcal{L}=-\frac{1}{4}F_{\alpha\nu}F^{\alpha\nu}-\frac{1}{4}W_{\alpha\nu
\rho\varphi}F^{\alpha\nu}F^{\rho\varphi}-J_{\alpha}A^{\alpha},
\end{equation}
where the background tensor $W_{\alpha\nu\rho\varphi}$ has the same symmetries
as the Riemann tensor $\left[  W_{\alpha\nu\rho\varphi}=-W_{\nu\alpha
\rho\varphi},\text{ }W_{\alpha\nu\rho\varphi}=-W_{\alpha\nu\varphi\rho},\text{
}W_{\alpha\nu\rho\varphi}=W_{\rho\varphi\alpha\nu}\right]  $ and a double null
trace, $W^{\alpha\beta}{}_{\alpha\beta}=0,$ implying 19 components. This
tensor $W_{\alpha\nu\rho\varphi}$ can be written in terms of four $3\times3$
matrices $\kappa_{DE},\kappa_{DB},\kappa_{HE},\kappa_{HB},$ defined in Refs.
\cite{KM1,KM2} as
\begin{equation}
\left(  \kappa_{DE}\right)  ^{jk}=-2W^{0j0k},\left(  \kappa_{HB}\right)
^{jk}=\frac{1}{2}\epsilon^{jpq}\epsilon^{klm}W^{pqlm},\left(  \kappa
_{DB}\right)  ^{jk}=-\left(  \kappa_{HE}\right)  ^{kj}=\epsilon^{kpq}W^{0jpq}.
\end{equation}
The matrices $\kappa_{DE},\kappa_{HB}$ contain the parity-even components and
possess together eleven independent components, while $\kappa_{DB},\kappa
_{HE}$ possess together eight components and describe the parity-odd sector of
$W_{\alpha\nu\rho\varphi}.$ Four tilde matrices and one trace element can
be\textbf{ }written as suitable combinations of $\kappa_{DE},\kappa
_{DB},\kappa_{HE},\kappa_{HB}$:\textbf{ }%
\begin{align}
\left(  \widetilde{\kappa}_{e+}\right)  ^{jk}  &  =\frac{1}{2}(\kappa
_{DE}+\kappa_{HB})^{jk},\text{ }\left(  \widetilde{\kappa}_{e-}\right)
^{jk}=\frac{1}{2}(\kappa_{DE}-\kappa_{HB})^{jk}-\frac{1}{3}\delta^{jk}%
(\kappa_{DE})^{ii},\\
\left(  \widetilde{\kappa}_{o+}\right)  ^{jk}  &  =\frac{1}{2}(\kappa
_{DB}+\kappa_{HE})^{jk},\text{ }\left(  \widetilde{\kappa}_{o-}\right)
^{jk}=\frac{1}{2}(\kappa_{DB}-\kappa_{HE})^{jk},\text{ }\widetilde{\kappa
}_{tr}=\frac{1}{3}(\kappa_{DE})^{ii}.
\end{align}

From the eleven independent components of the matrices $\widetilde{\kappa
}_{e+},\widetilde{\kappa}_{e-}$, the five\ elements enclosed in $\widetilde
{\kappa}_{e+}$ are constrained by birefringence to the level of 1 part in
$10^{32}$ (see Refs. \cite{KM1,KM2}), there remaining six non-birefringent
ones (the trace element and the five components of the matrix $\widetilde
{\kappa}_{e-})$ to be constrained by other methods. From the eight elements of
the parity-odd sector, five (contained in the matrix $\widetilde{\kappa}%
_{o-})$ are tightly bounded by birefringence, there remaining only three
components (belonging to $\widetilde{\kappa}_{o+})$, which were parameterized
as the $\kappa$ vector \cite{Kob}, written as $\kappa^{j}=\frac{1}{2}%
\epsilon^{jpq}\left(  \kappa_{DB}\right)  ^{pq}$. In some recent papers
\cite{Klink}, the absence of Cerenkov radiation from ultrahigh-energy cosmic
rays (UHECRs) has been used to state bounds at the level of 1 part in
$10^{18}$ on the nine nonbirefringent terms of $W_{\alpha\nu\rho\varphi},$
belonging both to the parity-even and parity-odd sectors.

In Ref. \cite{Manojr}, there was performed an analysis focused on the three
non-birefringent components ($\kappa^{j}$) of the parity-odd sector of $W$
(the parity-even components were taken as null in order to isolate the
parity-odd sector physics). The stationary classical solutions for the Maxwell
electrodynamics modified by these three LIV\ coefficients were properly
evaluated by means of the Green's method. With these solutions, it was
described a device able to yield a nice upper bound, $\kappa^{j}<10^{-16},$ in
the context of an Earth-based experiment.

The aim of the present work is to study the stationary aspects of the
classical electrodynamics stemming from the parity-even sector of the tensor
$W_{\alpha\nu\rho\varphi}.$ For that, we use the following parameterization
$\kappa_{DE}=-\kappa_{HB},$ \ and we consider as null the parity-odd sector.
The goal is to determine how the the six non-birefringent parity-even
components modify the classical and stationary solutions for Maxwell
electromagnetism. Certainly, the idea is also to use the results obtained to
properly constrain the magnitude of the LIV coefficients.

This work is organized as follows. In Sec. II, we write the wave equations and
modified Maxwell equations and apply the Green method in order to obtain the
required stationary solutions. In Sec. III, we present our final remarks and
describe a measurement device able to yielding an upper bound on the
\textbf{LIV} parameter as stringent as $\left(  \widetilde{\kappa}%
_{e-}\right)  ^{ij}<2.9\times10^{-20}$.

\section{Wave equations and stationary classical solutions}

Here, the focus is on the classical properties of the parity-even part of the
tensor $W$, particularly on those produced by the six non-birefringent
components (located in the matrix $\widetilde{\kappa}_{e-}$ plus the trace
element \textbf{$\widetilde{\kappa}_{\text{tr}}).$} Hence, we take the
parity-odd sector as null $\left(  \kappa_{DB}=\kappa_{HE}=0\right)  $ to
isolate the physics of the even sector. Further, we adopt the following
parameterization $\left(  \kappa_{DE}\right)  =-\left(  \kappa_{HB}\right)  ,$
which implies $\left(  \widetilde{\kappa}_{e+}\right)  =0,$ by considering the
stringent bound imposed by birefringence data \cite{KM1,KM2}. Moreover, it
implies the following relation for the non-birefringent components:
\begin{equation}
\text{ }\left(  \widetilde{\kappa}_{e-}\right)  ^{jk}=(\kappa_{DE}%
)^{jk}-n\delta^{jk},\text{ }n=\frac{1}{3}\delta^{jk}Tr(\kappa_{DE}).
\end{equation}

In order to evaluate the classical solutions of this model, we write the wave
equation for the four-potential%
\begin{equation}
\square A^{\alpha}-2W^{\alpha\nu\rho\lambda}\partial_{\nu}\partial_{\rho
}A_{\lambda}=J^{\alpha},
\end{equation}
which yields two\ differential equations, one for the scalar potential, and
one for the vector potential,
\begin{align}
\left[  (1+n)\partial_{t}^{2}-(1+n)\nabla^{2}-\left(  \widetilde{\kappa}%
_{e-}\right)  ^{ij}\partial_{i}\partial_{j}\right]  A_{0}+\left(
\widetilde{\kappa}_{e-}\right)  ^{ij}\partial_{i}\partial_{t}A_{j}  &
=\rho,\\
\left[  (1+n)\partial_{t}^{2}-(1-n)\nabla^{2}\right]  A_{i}-2n\partial
_{t}\partial_{i}A_{0}-\left(  \widetilde{\kappa}_{e-}\right)  ^{ij}%
\partial_{t}E_{j}-\epsilon_{ipj}\left(  \widetilde{\kappa}_{e-}\right)
^{jl}\partial_{p}B_{l}  &  =j_{i}.
\end{align}
where we have used\textbf{ }$E_{j}=-F_{0j},$ $B_{i}=\frac{1}{2}\epsilon
_{ipj}F_{pj},$ $\left(  \widetilde{\kappa}_{e-}\right)  ^{ij}=\left(
\widetilde{\kappa}_{e-}\right)  _{ij}.$ At the stationary regime, such
equations are read%
\begin{align}
\left[  (1+n)\nabla^{2}+\left(  \widetilde{\kappa}_{e-}\right)  ^{ij}%
\partial_{i}\partial_{j}\right]  A_{0}  &  =-\rho,\label{A_zero1}\\
\left[  (1-n)\nabla^{2}\right]  A_{i}+\epsilon_{ipj}\left(  \widetilde{\kappa
}_{e-}\right)  ^{jl}\partial_{p}B_{l}  &  =-j_{i}. \label{A_vector1}%
\end{align}

These equations reveal that the electric and magnetic sectors are decoupled
(in the stationary regime) in contrast with the electrodynamics of the
parity-odd sector, in which these sectors are entirely entwined (see Ref.
\cite{Manojr}). Applying the differential operator $\epsilon_{abi}%
\partial_{b\text{ }}$to Eq. (\ref{A_vector1}), we obtain the following
differential equation for the magnetic field%
\begin{equation}
\left[  \left(  (1-n)\delta_{al}-\left(  \widetilde{\kappa}_{e-}\right)
^{al}\right)  \nabla^{2}+\left(  \widetilde{\kappa}_{e-}\right)  ^{jl}%
\partial_{a}\partial_{j}\right]  B_{l}=-\left(  \nabla\times j\right)
_{a}.\label{BW1}%
\end{equation}
While the homogeneous Maxwell equations remain unmodified ($~\nabla
\times\mathbf{E}+\partial_{t}\mathbf{B}$ $=0\mathbf{,}$ $\nabla\mathbf{\cdot
B}=0)$, the inhomogeneous\textbf{ }ones (Gauss and Ampere law) are altered,
taking the form
\begin{align}
(1+n)\nabla\mathbf{\cdot E}-\left(  \widetilde{\kappa}_{e-}\right)
^{ij}\partial_{i}E_{j} &  =\rho,\\
(1+n)\partial_{t}E_{i}-(1-n)\left(  \nabla\times B\right)  _{i}+\epsilon
_{ijr}\left(  \widetilde{\kappa}_{e-}\right)  ^{rl}\partial_{j}B_{l}+\left(
\widetilde{\kappa}_{e-}\right)  ^{iq}\partial_{t}E_{q} &  \mathbf{=-}j_{i}.
\end{align}
In the stationary regime, the latter equation provides%
\begin{equation}
(1-n)\left(  \nabla\times B\right)  _{i}-\epsilon_{ijr}\left(  \widetilde
{\kappa}_{e-}\right)  ^{rl}\partial_{j}B_{l}\mathbf{=}j_{i},
\end{equation}
which under the action of the operator curl operator ($\epsilon_{abi}%
\partial_{b})$ yields the same expression as Eq. (\ref{BW1}).

\subsection{The Green's function for the scalar potential}

The solution for the scalar potential may be obtained by the Green's method.
The Green's function for Eq.(\ref{A_zero1}) fulfills
\begin{equation}
\left[  (1+n)\nabla^{2}+\left(  \widetilde{\kappa}_{e-}\right)  _{ij}%
\partial_{i}\partial_{j}\right]  G(\mathbf{r}-\mathbf{r}^{\prime})=\delta
^{3}(\mathbf{r}-\mathbf{r}^{\prime}),
\end{equation}
and the scalar potential is given as
\begin{equation}
A_{0}\left(  \mathbf{r}\right)  =-\int G(\mathbf{r-r}^{\prime})\rho
(\mathbf{r}^{\prime})d^{3}\mathbf{r}^{\prime}. \label{A0S}%
\end{equation}

The Green's function in Fourier space is given as $G(r-r^{\prime
})=\displaystyle\left(  2\pi\right)  ^{-3}\int d^{3}p~\tilde{G}\left(
\mathbf{p}\right)  \exp\left[  -i\mathbf{p}\cdot(\mathbf{r}-\mathbf{r}%
^{\prime})\right]  $, so that we obtain
\begin{equation}
\tilde{G}\left(  \mathbf{p}\right)  \simeq-\frac{1}{\mathbf{p}^{2}}\left[
1-n-\left(  \widetilde{\kappa}_{e-}\right)  ^{ij}\frac{p_{i}p_{j}}%
{\mathbf{p}^{2}}\right]  ,
\end{equation}
at first order in the LIV parameters. Remembering that the LIV coefficients
are small, we used~$\left[  1+n+\left(  \widetilde{\kappa}_{e-}\right)
^{ij}p_{i}p_{j}/\mathbf{p}^{2}\right]  ^{-1}\simeq\left[  1-n-\left(
\widetilde{\kappa}_{e-}\right)  ^{ij}p_{i}p_{j}/\mathbf{p}^{2}\right]  $.
Carrying out the inverse Fourier transform, the Green's function takes the
following form:
\begin{equation}
G(\mathbf{r}-\mathbf{r}^{\prime})=-\frac{1}{4\pi}\left\{  \left(  1-n\right)
\frac{1}{\left\vert \mathbf{r-r}^{\prime}\right\vert }+\frac{\left(
\widetilde{\kappa}_{e-}\right)  ^{ij}\mathbf{(r-r}^{\prime})_{i}%
\mathbf{(r-r}^{\prime})_{j}}{2\left\vert \mathbf{r-r}^{\prime}\right\vert
^{3}}\right\}  \label{Green1}%
\end{equation}
It presents a genuine Coulomb contribution screened by the factor $\left(
1-n\right)  $ and a non-Coulomb contribution related to the LIV non-isotropic
coefficients $\left(  \widetilde{\kappa}_{e-}\right)  _{ij}$. The overall
behavior, $r^{-1}$, remains the same as happens in the Maxwell electrodynamics.

Using the Green function (\ref{Green1}) and Eq. (\ref{A0S}), the scalar
potential due to\ a general charge distribution $\left[  \rho\left(
\mathbf{r}^{\prime}\right)  \right]  $ is
\begin{equation}
A_{0}\left(  \mathbf{r}\right)  =\frac{1}{4\pi}\left\{  \left(  1-n\right)
\int d^{3}\mathbf{r}^{\prime}~\frac{\rho\left(  \mathbf{r}^{\prime}\right)
}{\left\vert \mathbf{r-r}^{\prime}\right\vert }+\left(  \widetilde{\kappa
}_{e-}\right)  ^{ij}\int d^{3}\mathbf{r}^{\prime}~\frac{\mathbf{(r-r}^{\prime
})_{i}\mathbf{(r-r}^{\prime})_{j}}{2\left\vert \mathbf{r-r}^{\prime
}\right\vert ^{3}}\rho\left(  \mathbf{r}^{\prime}\right)  \right\}
,\ \label{Scalar}%
\end{equation}
which implies the following electric field strength:
\begin{align}
\mathbf{E}^{i}\left(  \mathbf{r}\right)   &  =\frac{1}{4\pi}\biggl\{\left(
1-n\right)  \int d^{3}\mathbf{r}^{\prime}~\rho\left(  \mathbf{r}^{\prime
}\right)  \frac{\left(  \mathbf{r-r}^{\prime}\right)  ^{i}}{\left\vert
\mathbf{r-r}^{\prime}\right\vert ^{3}}-\left(  \widetilde{\kappa}_{e-}\right)
^{ij}\int d^{3}\mathbf{r}^{\prime}\rho\left(  \mathbf{r}^{\prime}\right)
~\frac{\mathbf{(r-r}^{\prime})_{j}}{\left\vert \mathbf{r-r}^{\prime
}\right\vert ^{3}}\nonumber\\
&  +3\left(  \widetilde{\kappa}_{e-}\right)  ^{lj}\int d^{3}\mathbf{r}%
^{\prime}\rho\left(  \mathbf{r}^{\prime}\right)  \frac{\left(  \mathbf{r-r}%
^{\prime}\right)  _{l}\left(  \mathbf{r-r}^{\prime}\right)  _{j}\left(
\mathbf{r-r}^{\prime}\right)  ^{i}}{2\left\vert \mathbf{r-r}^{\prime
}\right\vert ^{5}}\biggr\},\ \label{Elect}%
\end{align}

With this expression, we may immediately evaluate the scalar potential and
electric field strength for a point-like charge at rest $\left[
\rho(\mathbf{r}^{\prime})=q\delta(\mathbf{r}^{\prime})\right]  $, yielding
\begin{align}
A_{0}\left(  \mathbf{r}\right)   &  =\frac{q}{4\pi}\left\{  \left(
1-n\right)  \frac{1}{r}+\left(  \widetilde{\kappa}_{e-}\right)  ^{ij}%
\frac{r_{i}r_{j}}{2r^{3}}\right\}  ,\\
\mathbf{E}^{i}\left(  \mathbf{r}\right)   &  =\frac{q}{4\pi}\left\{  \left(
1-n+\frac{3}{2}\left(  \widetilde{\kappa}_{e-}\right)  ^{lj}\frac{r_{l}r_{j}%
}{r^{2}}\right)  \frac{r^{i}}{r^{3}}-\left(  \widetilde{\kappa}_{e-}\right)
^{ij}~\frac{r_{j}}{r^{3}}\right\}  .
\end{align}
The scalar potential and the electric field present a genuine Coulomb
contribution, with the screening factor $\left(  1-n\right)  ,$ and a
non-Coulomb contribution related to the LIV coefficient $\left(
\widetilde{\kappa}_{e-}\right)  ^{ij}.$ This latter term leads to variations
of the scalar potential and electric field along a circular path around the
point-like charge. Such an effect can be used to impose an upper bound on the
LIV parameters, as will be described in Sec. III.

From the expressions (\ref{Scalar}) and (\ref{Elect}), for an arbitrary charge
distribution, we express the scalar potential and the electric field in the
dipolar approximation ($\left\vert \mathbf{r-r}^{\prime}\right\vert
^{-1}=r^{-1}+\mathbf{r\cdot r}^{\prime}/r^{3})$:%
\begin{align}
A_{0}\left(  \mathbf{r}\right)   &  =\frac{1}{4\pi}\left\{  \left(
1-n\right)  ~\left[  \frac{q}{r}+\frac{\mathbf{r\cdot P}_{e}}{r^{3}}\right]
+\frac{\left(  \widetilde{\kappa}_{e-}\right)  ^{ij}}{2r^{3}}\left[
r_{i}r_{j}(q+3\frac{\mathbf{r\cdot P}_{e}}{r^{2}})-r_{i}P_{ej}-r_{j}%
P_{ei}\right]  \right\}  ,\ \\
\mathbf{E}^{i}\left(  \mathbf{r}\right)   &  =\frac{1}{4\pi}\biggl\{\left(
1-n\right)  \left[  \frac{q}{r^{3}}r^{i}-\left(  \frac{P_{e}^{i}}{r^{3}}%
-\frac{3\left(  \mathbf{r}\cdot\mathbf{P}_{e}\right)  }{r^{5}}r^{i}\right)
\right]  -\left(  \widetilde{\kappa}_{e-}\right)  ^{ij}\frac{~1}{r^{3}}\left(
qr_{j}-P_{ej}+3\frac{\left(  \mathbf{r}\cdot\mathbf{P}_{e}\right)  }{r^{2}%
}r_{j}\right) \nonumber\\
&  +3\left(  \widetilde{\kappa}_{e-}\right)  ^{lj}\frac{r_{l}r_{j}}{2r^{5}%
}\left[  qr^{i}-P_{e}^{i}+5\frac{\left(  \mathbf{r}\cdot\mathbf{P}_{e}\right)
}{r^{2}}r^{i}\right]  -3\left(  \widetilde{\kappa}_{e-}\right)  ^{lj}%
\frac{r_{l}P_{ej}}{r^{5}}r^{i}\biggr\},
\end{align}
where \textbf{$q=\int\,\rho(\mathbf{r}^{\prime})d^{3}\mathbf{r}^{\prime}$
}and\textbf{ } $\mathbf{P}_{e}\mathbf{=}\int\mathbf{r}^{\prime}\rho
(\mathbf{r}^{\prime})d^{3}\mathbf{r}^{\prime}$ is the electric dipole moment.
The LIV terms in $\left(  \widetilde{\kappa}_{e-}\right)  ^{lj}$ break the
radial symmetry giving a non-Coulomb behavior to the static solutions. Despite
the large number of terms in these solutions, we verify that the electric
field preserves the $r^{-2}$ and the $r^{-3}$ decaying behaviors for the
monopole and dipole moments, respectively, as occurs in the pure Maxwell
electrodynamics. Obviously, it is a consequence of the dimensionless character
of the LIV coefficients.

\subsection{The Green's function for the magnetic field}

Now, we search for an explicit solution for the magnetic field. The Green's
function for the magnetic field equation of motion (\ref{BW1}) is written as%

\begin{equation}
\left[  \left(  (1-n)\delta_{al}-\left(  \widetilde{\kappa}_{e-}\right)
^{al}\right)  \nabla^{2}+\left(  \widetilde{\kappa}_{e-}\right)  ^{jl}%
\partial_{a}\partial_{j}\right]  G_{lb}\left(  \mathbf{r-r}^{\prime}\right)
=\delta_{ab}\delta^{3}(\mathbf{r}-\mathbf{r}^{\prime}).
\end{equation}
Using the Fourier representation and having much care in the tensor inversion
procedure, we obtain in the momentum space
\begin{equation}
\tilde{G}_{ab}\left(  \mathbf{p}\right)  =-\frac{1}{\mathbf{p}^{2}}\left[
\left(  1+n\right)  \delta_{ab}+\left(  \widetilde{\kappa}_{e-}\right)
_{ab}-\left(  \widetilde{\kappa}_{e-}\right)  _{cb}\frac{p_{a}p_{c}%
}{\mathbf{p}^{2}}\right]  .
\end{equation}
Performing the inverse Fourier transformation, we attain the following
expression:
\begin{equation}
G_{ab}(\mathbf{r}-\mathbf{r}^{\prime})=-\frac{1}{4\pi}\frac{1}{\left\vert
\mathbf{r-r}^{\prime}\right\vert }\left\{  \left(  1+n\right)  \delta
_{ab}+\frac{\left(  \widetilde{\kappa}_{e-}\right)  _{ab}}{2}+\frac{\left(
\widetilde{\kappa}_{e-}\right)  _{cb}\mathbf{(r-r}^{\prime})_{a}%
\mathbf{(r-r}^{\prime})_{c}}{2\left\vert \mathbf{r-r}^{\prime}\right\vert
^{2}}\right\}  ,
\end{equation}
with which the magnetic field is then written as%
\begin{equation}
B^{i}\left(  \mathbf{r}\right)  =-\int d^{3}\mathbf{r}^{\prime}~G_{ij}\left(
\mathbf{r}-\mathbf{r}^{\prime}\right)  \left(  \nabla^{\prime}\times
\mathbf{J}\left(  \mathbf{r}^{\prime}\right)  \right)  ^{j}.
\end{equation}
It leads to the explicit solution:
\begin{equation}
B^{i}\left(  \mathbf{r}\right)  =\frac{1}{4\pi}\left\{  \left[  \left(
1+n\right)  \delta_{ib}+\frac{1}{2}\left(  \widetilde{\kappa}_{e-}\right)
_{ib}\right]  \int d^{3}\mathbf{r}^{\prime}~\frac{\left(  \nabla\times
j\left(  \mathbf{r}^{\prime}\right)  \right)  ^{b}}{\left\vert \mathbf{r-r}%
^{\prime}\right\vert }+\frac{\left(  \widetilde{\kappa}_{e-}\right)  ^{lj}}%
{2}\int d^{3}\mathbf{r}^{\prime}\left[  ~\frac{\left(  \nabla\times j\left(
\mathbf{r}^{\prime}\right)  \right)  ^{l}}{\left\vert \mathbf{r-r}^{\prime
}\right\vert ^{3}}\mathbf{(r-r}^{\prime})_{j}(\mathbf{r-r}^{\prime}%
)^{i}\right]  \right\}  . \label{BS1}%
\end{equation}
After a certain algebraic effort, a dipolar expansion for the magnetic field
is achieved as well, yielding%
\begin{equation}
B^{i}\left(  \mathbf{r}\right)  =\frac{1}{4\pi}\biggl\{\left(  1+n\right)
\left(  -\frac{m^{i}}{r^{3}}+\frac{3\left(  \mathbf{r\cdot m}\right)  }{r^{5}%
}r^{i}\right)  -\left(  \widetilde{\kappa}_{e-}\right)  ^{ib}\frac
{\mathbf{m}_{b}}{r^{3}}-\left(  \kappa_{e-}\right)  _{pb}r_{p}r_{b}\left[
\frac{3~}{2}\frac{m^{i}}{r^{5}}-\frac{15~}{2}\frac{\left(  \mathbf{r\cdot
m}\right)  }{r^{7}}r^{i}\right]  \biggr\} \label{BS2b}%
\end{equation}
where we have considered a localized and divergenceless current density
distribution\textbf{ $\mathbf{J}$, }and\textbf{ $\mathbf{m=}\frac{1}{2}%
\int\mathbf{r^{\prime}}\times\mathbf{J}(\mathbf{r^{\prime}})d^{3}%
\mathbf{r^{\prime}}$ }is the magnetic dipole moment. In Eq. (\ref{BS2b}) the
first term inside the parentheses is the usual Maxwell contribution, just
corrected by the $\left(  1+n\right)  $ factor. The terms that are
proportional to the LIV coefficients,\textbf{ }$\left(  \widetilde{\kappa
}_{e-}\right)  ^{ib}$\textbf{, }ascribe to the solution a directional
dependence or anisotropic character. In principle, such a directional
dependence could be used to impose an upper bound on the LIV parameters. In
Sec. III, we show that the attained bound is not as restrictive as desired.

\section{Final remarks}

We should now compare these parity-even stationary solutions with the
parity-odd ones derived in Ref. \cite{Manojr}. At the stationary regime, the
main difference is that now the electric and magnetic sectors are not coupled
by the LIV tensor anymore. In the parity-odd case, a stationary current is
able to produce an electric field as much as a static charge can generate a
magnetic field. As such an interconnection does not appear in the present
case, the manifestation of pure LIV electromagnetic effects (aside from
Maxwell ones), as the production of magnetic field by a static point-like
charge (see Ref.\cite{Manojr}), are absent. Now, the LIV effects appear as
small corrections for the\ usual Maxwell's electric and magnetic fields.\ Yet,
the LIV effects can still be identified by means of suitable devices, as it is
discussed below. Apart from this difference, the solutions of the parity-even
and parity-odd sectors possess some similarities. Indeed, the electric field
for a point-like charge (in both sectors) exhibits an asymptotic behavior as
$r^{-2}$, while a stationary current provides a magnetic field whose dipolar
expansion is proportional to $mr^{-3}$. This is ascribed to the dimensionless
character of the tensor $W$.

The attained magnetic field solution does not lead to good upper bounds on the
magnitude of the parameters $n$ and $\left(  \widetilde{\kappa}_{e-}\right)
_{ib}$ when we take as reference the Earth's magnetic field. In fact,
proceeding in a similar way as in Ref. \cite{Jackiw}, we assert that the LIV
tensor must not imply a magnetic field contribution larger than $10^{-4}$G
(otherwise it would be detected). From Eq. (\ref{BS2b}), we observe that the
LIV terms are always proportional to $(m/r^{3})$. Assuming that $m$ represents
the Earth's magnetic dipole, and $R_{\oplus}$ the Earth's radius, it holds the
following ratio $m/R_{\oplus}^{3}=0.3$ G (see Ref. \cite{Jackiw}). This
procedure, however, implies a non-restrictive bound: $n\leq10^{-4}.$

A much better bound for the parameters $\left(  \widetilde{\kappa}%
_{e-}\right)  ^{ij}$ can be attained from the expression for the scalar
potential. The idea is to evaluate the scalar potential generated by a charged
sphere in different outer points located at the same distance from the center
of the sphere, observing the difference of potential induced by the
non-Coulomb LIV term. The starting point is the expression for the potential
generated by a conducting sphere of radius $R$ and charge $q$ (uniformly
distributed over its surface), which can be achieved by replacing the charge
density for a sphere, $\rho\left(  \mathbf{r}^{\prime}\right)  =q\delta
(r^{\prime}-R)/(4\pi R^{2}),$ in Eq. (\ref{Scalar}). Using Fourier
integrations (see the appendix), the potential is (for $r>R)$%

\begin{equation}
A_{0}\left(  \mathbf{r}\right)  =\frac{1}{4\pi}\biggl\{(1-n)\frac{q}{r}%
+\frac{\left(  \widetilde{\kappa}_{e-}\right)  ^{ab}}{2}\left[  \frac
{r_{a}r_{b}\left(  r^{2}-R^{2}\right)  }{r^{5}}\right]  \biggr\}.
\end{equation}
We see that the term in $\left(  \widetilde{\kappa}_{e-}\right)  ^{ij}$ breaks
the radial symmetry of the potential, implying potential variations along a
circular path around the center. We now expand the term $\left(
\widetilde{\kappa}_{e-}\right)  ^{ab}r_{a}r_{b}$ at the form
\begin{align*}
\left(  \widetilde{\kappa}_{e-}\right)  ^{ab}r_{a}r_{b}  &  =\left(
\widetilde{\kappa}_{e-}\right)  ^{11}\left[  \left(  r_{1}\right)
^{2}-\left(  r_{3}\right)  ^{2}\right]  +\left(  \widetilde{\kappa}%
_{e-}\right)  ^{22}\left[  \left(  r_{2}\right)  ^{2}-\left(  r_{3}\right)
^{2}\right] \\
&  +2\left(  \widetilde{\kappa}_{e-}\right)  ^{12}r_{1}r_{2}+2\left(
\widetilde{\kappa}_{e-}\right)  ^{13}r_{1}r_{3}+2\left(  \widetilde{\kappa
}_{e-}\right)  ^{23}r_{2}r_{3},
\end{align*}
where we have used the traceless matrix%
\begin{equation}
\left(  \widetilde{\kappa}_{e-}\right)  =\left(
\begin{array}
[c]{ccc}%
\left(  \widetilde{\kappa}_{e-}\right)  _{11} & \left(  \widetilde{\kappa
}_{e-}\right)  _{12} & \left(  \widetilde{\kappa}_{e-}\right)  _{13}\\
\left(  \widetilde{\kappa}_{e-}\right)  _{12} & \left(  \widetilde{\kappa
}_{e-}\right)  _{22} & \left(  \widetilde{\kappa}_{e-}\right)  _{23}\\
\left(  \widetilde{\kappa}_{e-}\right)  _{13} & \left(  \widetilde{\kappa
}_{e-}\right)  _{23} & -\left(  \widetilde{\kappa}_{e-}\right)  _{11}-\left(
\widetilde{\kappa}_{e-}\right)  _{22}%
\end{array}
\right)  .
\end{equation}
Then, we can conceive of an experiment to measure the electrostatic potential
generated by a 1 C charged sphere of radius $R$ (maintained in vacuum) in two
distinct outer points, A and B, located at the a circle of radius $r>R$ on the
$x-y$ plane. We consider the points A and B symmetrically disposed in relation
to the $y$-axis at the positions: $A=r(\cos\phi,\sin\phi,0),$ $B=r(-\cos
\phi,\sin\phi,0)$. Then, the difference of potential between these points is
simply
\begin{equation}
\Delta A_{0}=A_{0}(A)-A_{0}(B)=\frac{q}{4\pi}\left(  \widetilde{\kappa}%
_{e-}\right)  ^{12}\sin2\phi~\frac{\left(  r^{2}-R^{2}\right)  }{r^{3}},
\end{equation}
for$\ r>R.$ For $\phi=\pi/4$ and a 1 C charge, such difference of potentials
equal to
\begin{equation}
\Delta A_{0}=9\times10^{9}\left(  \widetilde{\kappa}_{e-}\right)  ^{12}%
~\frac{\left(  r^{2}-R^{2}\right)  }{r^{3}}. \label{A08}%
\end{equation}
For attaining the best bound, we should consider the maximum value of Eq.
(\ref{A08}). So, it must be evaluated at the point $r=R\sqrt{3},$ in which the
expression $\left(  r^{2}-R^{2}\right)  r^{-3}$ has a maximum. For a charged
sphere of unitary radius $\left(  R=1m\right)  ,$ we obtain $\Delta
A_{0}=3.46\times10^{9}\left(  \widetilde{\kappa}_{e-}\right)  ^{12}~V.$ Given
the existence of sensitive methods for measurement of the potential able to
detect slight variations of 1 part in $10^{10}$ V, we can infer that the
voltage difference of Eq. (\ref{A08}) cannot be larger than $10^{-10}$ V, that
is, $3.46\times10^{9}\left(  \widetilde{\kappa}_{e-}\right)  ^{12}<10^{-10}.$
This condition leads to $\left(  \widetilde{\kappa}_{e-}\right)
^{12}<2.9\times10^{-20}.$ Choosing pairs of points on the planes $y-z$ and
$x-z,$ this upper limit holds equivalently for $\left(  \widetilde{\kappa
}_{e-}\right)  ^{23}$ and $\left(  \widetilde{\kappa}_{e-}\right)  ^{13}$.

This device can also be used to set up\ an upper bound on the diagonal
components $\left(  \widetilde{\kappa}_{e-}\right)  ^{ii}$. For constraining
$\left(  \widetilde{\kappa}_{e-}\right)  ^{11}$, we take the points A and B in
the positions: $A=r(1,0,0),$ $B=r(0,0,1).$ The difference of potential between
these points is
\begin{equation}
\Delta A_{0}=A_{0}(A)-A_{0}(B)=\frac{q}{4\pi}\left(  \widetilde{\kappa}%
_{e-}\right)  ^{11}~\left[  \frac{r^{2}-R^{2}}{r^{3}}\right]  ,
\end{equation}
which leads to the same bound obtained for the non-diagonal components:
$\left(  \widetilde{\kappa}_{e-}\right)  ^{11}<2.9\times10^{-20}.$ Choosing
two points on the $y-z$ plane$,$ $A=r(0,1,0),$ $B=r(0,0,1),$ this bound can be
stated to $\left(  \widetilde{\kappa}_{e-}\right)  ^{22}.$ Thus, we conclude
that by means of this experiment it is possible to establish an upper bound as
stringent as
\[
\left(  \widetilde{\kappa}_{e-}\right)  ^{ij}<2.9\times10^{-20},
\]
for the five non-isotropic components of the traceless matrix $\left(
\widetilde{\kappa}_{e-}\right)  ^{ij}.$ This is a nice bound for an
Earth-based experiment, as good as the best bounds stated from astrophysical
data analysis of UHECRs \cite{Klink}.

As the isotropic component $n$ does not break the spherical symmetry of the
potential, this kind of experiment does not provide any way for bounding it.
This component induces a slight screening on the Coulomb potential that may be
interpreted as a charge screening. An experiment able to constrain $n$ could
be based on a charge or potential screening measurement. In this case, the
major difficult is that the tiny LIV\ effect is disguised by the dominant
Maxwell's behavior, avoiding its isolation. So, the LIV\ effect stays limited
by the experimental imprecisions of the device. A two-sided bound was recently
stated for this coefficient in the context of quantum electrodynamics decay
processes modified by this LIV parameter \cite{Klink3}.

Finally, we should note that this work completes the calculation of the
stationary solutions of Maxwell's electromagnetism modified by the
non-birefringent elements of the abelian CPT-even and LIV sector of the
standard model extension, a task initiated in Refs. \cite{Bailey, Manojr}.

\appendix

\section{Evaluation of the scalar potential generated by a charged sphere}

By starting from Eq. (\ref{Scalar}), the scalar potential is rewritten as
\begin{equation}
A_{0}\left(  \mathbf{r}\right)  =\frac{1}{4\pi}\biggl\{(1-n)\int
d^{3}\mathbf{r}^{\prime}~\frac{\rho\left(  \mathbf{r}^{\prime}\right)
}{\left\vert \mathbf{r-r}^{\prime}\right\vert }-\frac{\left(  \widetilde
{\kappa}_{e-}\right)  ^{ab}}{2}\partial_{a}I_{b}\biggr\},\
\end{equation}
where
\begin{equation}
I_{b}=\int d^{3}\mathbf{r}^{\prime}\frac{\mathbf{(r-r}^{\prime})_{b}%
}{\left\vert \mathbf{r-r}^{\prime}\right\vert }\rho\left(  \mathbf{r}^{\prime
}\right)  , \label{IBB}%
\end{equation}
and using $\left(  \widetilde{\kappa}_{e-}\right)  _{ii}=\text{tr}\left(
\widetilde{\kappa}_{e-}\right)  =0$, we show that
\[
\left(  \widetilde{\kappa}_{e-}\right)  ^{ab}\frac{\mathbf{(r-r}^{\prime}%
)_{a}\mathbf{(r-r}^{\prime})_{b}}{\left\vert \mathbf{r-r}^{\prime}\right\vert
^{3}}=-\left(  \widetilde{\kappa}_{e-}\right)  ^{ab}\partial_{a}\left[
\frac{\mathbf{(r-r}^{\prime})_{b}}{\left\vert \mathbf{r-r}^{\prime}\right\vert
}\right]  .
\]
Knowing that the charge density for a charged sphere of radius $R$ is
$\rho\left(  \mathbf{r}^{\prime}\right)  =q\delta\left(  r^{\prime}-R\right)
/(4\pi R^{2})$, its Fourier transform is
\begin{equation}
\tilde{\rho}\left(  \mathbf{p}\right)  =\int d^{3}\mathbf{r}^{\prime
}~e^{i\mathbf{p\cdot r}\prime}\rho\left(  \mathbf{r}^{\prime}\right)
=q\frac{\sin\left(  pR\right)  }{pR}, \label{fresf}%
\end{equation}
with $p=|\mathbf{p|.}$ For evaluating the integral $I_{b}$, we use the Fourier
representation of the Coulomb potential
\begin{equation}
\frac{1}{\left\vert \mathbf{r-r}^{\prime}\right\vert }=4\pi\int\frac
{d^{3}\mathbf{p}}{\left(  2\pi\right)  ^{3}}\frac{e^{-i\mathbf{p\cdot}\left(
\mathbf{r-r}\prime\right)  }}{\mathbf{p}^{2}}.
\end{equation}
By substituting in the Eq. (\ref{IBB}) and using (\ref{fresf}) we obtain,
after some algebraic manipulations, the following expression for $I_{b}$:
\begin{align}
I_{b}  &  =4\pi r_{b}\frac{q}{R}\int\frac{d^{3}\mathbf{p}}{\left(
2\pi\right)  ^{3}}\left(  \frac{1}{\mathbf{p}^{2}}e^{-i\mathbf{p\cdot r}}%
\frac{\sin\left(  pR\right)  }{p}\right)  +4\pi\frac{q}{R}\partial_{b}%
\int\frac{d^{3}\mathbf{p}}{\left(  2\pi\right)  ^{3}}\left(  \frac
{1}{\mathbf{p}^{2}}e^{-i\mathbf{p\cdot r}}\frac{\sin\left(  pR\right)  }%
{p^{3}}\right) \nonumber\\
&  -4\pi q\partial_{b}\int\frac{d^{3}\mathbf{p}}{\left(  2\pi\right)  ^{3}%
}\left(  \frac{1}{\mathbf{p}^{2}}e^{-i\mathbf{p\cdot r}}\frac{\cos\left(
pR\right)  }{p^{2}}\right)  .
\end{align}
Solving these integrals, we obtain%
\begin{equation}
I_{b}=\int d^{3}\mathbf{r}^{\prime}\frac{\mathbf{(r-r}^{\prime})_{b}%
}{\left\vert \mathbf{r-r}^{\prime}\right\vert }\rho\left(  \mathbf{r}^{\prime
}\right)  =\frac{q}{r}r_{b}-\frac{qR^{2}}{3r^{3}}r_{b}.
\end{equation}

Thus, finally, we obtain the scalar potential generated by the charged sphere
of radius $R$ for\textbf{ ($r>R$)}
\begin{equation}
A_{0}\left(  \mathbf{r}\right)  =\frac{1}{4\pi}\biggl\{(1-n)\frac{q}{r}%
+\frac{q\left(  \widetilde{\kappa}_{e-}\right)  ^{ab}}{2}\left[  \frac
{r_{a}r_{b}\left(  r^{2}-R^{2}\right)  }{r^{5}}\right]  \biggr\}.
\label{pot_esfera}%
\end{equation}

\begin{acknowledgments}
The authors are grateful to FAPEMA (Funda\c{c}\~{a}o de Amparo \`{a} Pesquisa
do Estado do Maranh\~{a}o), to CNPq (Conselho Nacional de Desenvolvimento
Cient\'{\i}fico e Tecnol\'{o}gico), and CAPES for financial support.
\end{acknowledgments}

\end{document}